\documentclass[twocolumn,showpacs,aps,pre,preprintnumbers,floatfix,amsmath,amssymb]{revtex4-1}

\usepackage{amsmath, graphicx, subfigure, bm, url, rotating, color,
  epsfig, dcolumn, multirow, setspace}
\usepackage[flushleft]{threeparttable}
\usepackage{natbib,hyperref}

\hypersetup{
  colorlinks   = true, 
  urlcolor     = red, 
  linkcolor    = red, 
  citecolor   = red 
}

\begin{document}

\title{First-principles investigation of organic photovoltaic materials C$_{60}$, C$_{70}$, [C$_{60}$]PCBM, and bis-[C$_{60}$]PCBM using a many-body $G_0W_0$-Lanczos approach}

\author{Xiaofeng Qian$^{1,2}$, Paolo Umari$^{3,4}$, and Nicola Marzari$^{1,5}$}

\affiliation{$^1$Department of Materials Science and Engineering,
  Massachusetts Institute of Technology, Cambridge, Massachusetts
  02139, USA}

\affiliation{$^2$Department of Materials Science and Engineering, Texas A\&M University, College Station, Texas 77843, USA}

\affiliation{$^3$Dipartimento di Fisica e Astronomia, Universitˆ di Padova, via Marzolo 8, I-35131 Padova, Italy}

\affiliation{$^4$CNR-IOM DEMOCRITOS, Theory@Elettra Group, c/o Sincrotrone Trieste, Area Science Park, Basovizza, I-34012 Trieste, Italy}

\affiliation{$^5$Theory and Simulations of Materials (THEOS), and National Center for Computational Design and Discovery of Novel Materials (MARVEL), \'Ecole Polytechnique F\'ed\'erale de Lausanne, 1015 Lausanne, Switzerland}

\date{\today}

\begin{abstract}
  
  We present a first-principles investigation of the excited-state
  properties of electron acceptors in organic photovoltaics including C$_{60}$, C$_{70}$,
  [6,6]-phenyl-C$_{61}$-butyric-acid-methyl-ester ([C$_{60}$]PCBM), and
  bis-[C$_{60}$]PCBM using many-body perturbation theory within the Hedin's
  $G_0W_0$ approximation and an efficient Lanczos approach.  Calculated vertical
  ionization potentials (VIP) and vertical electron affinities (VEA)
  of C$_{60}$ and C$_{70}$ agree very well with experimental values
  measured in gas phase. The density of states of all three
  molecules is also compared to photoemission and inverse
  photoemission spectra measured on thin-films, exhibiting a close
  agreement -- a rigid energy-gap renormalization owing to
  intermolecular interactions in the thin-films. In addition, it is
  shown that the low-lying unoccupied states of [C$_{60}$]PCBM are all derived
  from the highest-occupied molecular orbitals and the
  lowest-unoccupied molecular orbitals of fullerene C$_{60}$. The
  functional side group in [C$_{60}$]PCBM introduces a
  slight electron transfer to the fullerene cage, resulting in small
  decreases of both VIP and VEA. This small change of VEA provides a
  solid justification for the increase of open-circuit voltage when
  replacing fullerene C$_{60}$ with [C$_{60}$]PCBM as the electron acceptor
  in bulk heterojunction polymer solar cells.
    
\end{abstract}


\maketitle 

\section{introduction}

Organic photovoltaics (OPV), especially bulk heterojunction (BHJ)
type \cite{Tang86, YuGHWH95}, are becoming a very promising alternative to
the traditional silicon solar cell technology since the former can
provide renewable, sustainable, and low-cost clean
energy \cite{BrabecSH01, Brabec04, HoppeS04, GunesNS07, MayerSHRM07,
  ThompsonF08, KroonLHBD08, DennlerSB09, DeibelD10}. The power
conversion efficiency of BHJ-OPV has greatly improved over the last
decade from 1\% to more than 9\% by tuning morphology and blending
ratio \cite{ShaheenBSPFH01,ChirvasePHD04, HoppeNWKHHMS04,
  vanDurenYLBSHJ04, MaYGLH05, LiSHYMEY05, HoppeS06, KimCTCNDBGMHR06,
  PeetKCMMHB07, AyznerWTTS08, VandewalGOBBVLCVM08, ChenHLY09},
reducing interfacial power losses \cite{KimKLLMGH06, IrwinBHCM08},
increasing the range of light absorption with tandem cell
architecture \cite{ShrotriyaWLYY06, KawanoINS06, HadipourdWKHTWJB06,
  GilotWJ07, KimLCMNDH07, AmeriDLB09}, optimizing energy levels,
carrier mobility and optical absorption of low-bandgap
polymers \cite{CoakleyM04, MuhlbacherSMZWGB06, BlouinML07, BlouinL08,
  BlouinMGWBNBDTL08, HouCZLY08, LiZ08, ChengYH09, ChenC09, FloresOV09,
  HeL09, ParkRBCCMMLLH09, LiXLLXWT09, LiangWFTSLY09, ZouGBNTL09,
  ChenHZLYYYWL09, Liang10} and fullerenes and their derivatives \cite{WienkKVKHvJ03, KooistraKKPVKH07, LenesWKVHB08, RossCGSRKPWBVHD09, ParkRBCCMMLLH09, Pfuetzner2009, Tromholt2009, GuilbertRBMKFPMSHN12}, and improving device structures \cite{Dou12, He12}. In the BHJ type of OPV, low-band gap polymers
not only serve as electron donors, but also play several important
roles in exciton generation upon light absorption, exciton migration
and recombination, and hole transport.
 Therefore, extensive efforts
have been made to enhance sun-light absorption in polymers and
increase open-circuit voltage (${\rm V}_{\mathrm  oc}$) by lowering the highest-occupied
molecular orbital (HOMO) of polymers.

In contrast, fullerenes (C$_{60}$, C$_{70}$) and their derivatives,
such as [6,6]-phenyl-C$_{61}$-butyric-acid-methyl-ester ([C$_{60}$]PCBM)
and bis-[C$_{60}$]PCBM are often used as electron acceptors in OPV due to
their large electron affinities and high electron mobilities. It is
known from experimental results that [C$_{60}$]PCBM not only improves the
solubility of pure fullerenes but also helps increase the open-circuit
voltage \cite{BrabecCMSFRSH01}. 

However, the fundamental mechanism of
how the functional side group of [C$_{60}$]PCBM affects the  ${\rm V}_{\mathrm  oc}$
has not been fully understood yet. In principle, the maximum attainable ${\rm V}_{\mathrm  oc}$
can be expressed as the energy difference between the LUMO
level of the electron acceptor and that of the HOMO of the electron donor.
It is therefore interesting to see if the theoretically predicted  LUMO levels for electron acceptors can 
account for the difference in ${\rm V}_{\mathrm  oc}$ reported in experiments.

 A few theoretical works, based on density-functional theory (DFT),
have been carried out in the past to look into the electronic structure of the
above electron acceptors \cite{ZhangHLZJZX08, AkaikeKYTNSOS08,
  KanaiAKSNKNOS09} leading to  a good description of their structural 
  properties. However, the evaluation through DFT of electronic properties such as
 quasiparticle (QP) energy levels yields results only in qualitative agreement with experiment.
High-level quantum chemistry approach and many-body perturbation theory are state-of-the-art methods which
can go beyond DFT and provide accurate predictions for
excited-state properties. In the case of QP energies in
charged excitations upon electron removal/addition, many-body perturbation theory within Hedin's {\it GW}
approximation \cite{Hedin65, HedinL69,Strinati1980,Strinati1982} is computationally less demanding.

\begin{figure}[thpb]\centering
  \includegraphics[width=0.8\columnwidth]{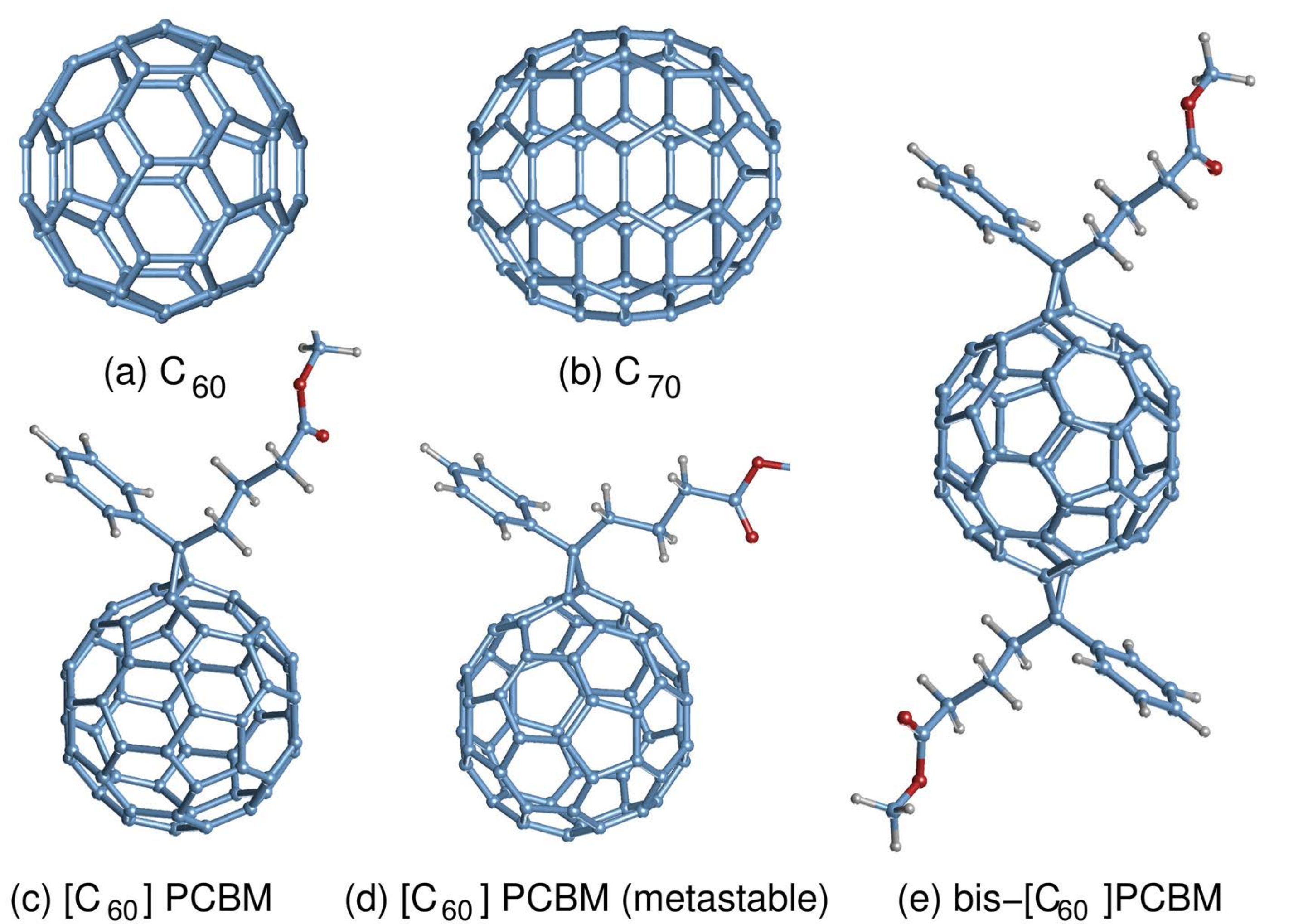}
  \caption{Structure of C$_{60}$, C$_{70}$, [C$_{60}$]PCBM, metastable-[C$_{60}$]PCBM, and
    bis-[C$_{60}$]PCBM.}
  \label{FIG:STRUCTURE}
\end{figure}

The {\it GW} method has already been applied to fullerenes in
some theoretical works \cite{TiagoKHR08, TiagoR09, BlaseAO11}. 
Here, we use the  recently
developed many-body {\it GW}-Lanczos approach which is particularly
effective in reaching numerical convergence \cite{UmariSB09,
  UmariSB10, StenuitCPFPGU10, UmariQMSGB11, QianUM11} in large
  atomic structures without suffering from
bottlenecks with respect to summing over a large number of empty
Kohn-Sham orbitals. This allows us to calculate the electronic structure of the electron acceptors including
 C$_{60}$, C$_{70}$, [C$_{60}$]PCBM, and bis-[C$_{60}$]PCBM at both the DFT and {\it GW} level, and compare them with experimental photoemission results. 
 Due to the difficulties in modeling extended and possibly nano-structured materials such as  [C$_{60}$]PCBM,
 we focus here on the isolated molecular limit,  addressing therefore changes in the position of LUMO levels
 rather than their actual values.
 Having validated the quality of our QP-energy levels we
 considered the differences in the maximum attainable ${\rm V}_\mathrm{oc}$, with 
 calculated ${\rm V}_\mathrm{oc}$ differences at the {\it GW} level found to be 
 in reasonable agreement with experiment.

\section{Methods}
\label{SEC:COMPUTATION}

Ground-state DFT calculations were performed 
using the {\it pw.x} code of the Quantum-ESPRESSO 
which is based on the planewave pseudopotentials scheme.
We used orthorhombic
supercells of 31.7$^3$, 31.7$^3$, 24$\times$19.2$\times$19.2, and
47.6$\times$34.3$\times$31.7 \AA$^3$ for C$_{60}$, C$_{70}$,
[C$_{60}$]PCBM, and bis-[C$_{60}$]PCBM, respectively, 
the generalized-gradient approximation (GGA) of exchange-correlation functional
in the Perdew-Burke-Ernzerhof (PBE) form \cite{PerdewBE96}, Troullier-Martins's norm-conserving
pseudopotentials, and a plane-wave basis set with a cutoff of 612.3 eV
for C$_{60}$ and C$_{70}$ and 816.3 eV for [C$_{60}$]PCBM, and
bis-[C$_{60}$]PCBM. Atomic structures were optimized with residual force
threshold of 0.026 eV/\AA, and are displayed in
Fig.~\ref{FIG:STRUCTURE}.  We considered not only [C$_{60}$]PCBM in its lowest energy configuration, but also a local-minimum metastable structure with an energy of 0.29 eV higher than that of the ground-state structure.

The {\it GW} QP-energy level $E_i$ for the  $i$-th orbital is  obtained
 in the so-called diagonal G$_0$W$_0$ scheme through the solution of the following
 self-consistent single-variable equation:
 \begin{equation}
 \label{eq:GW}
 E_i=\epsilon_i -\langle\psi_i|V_\mathrm{xc}|\psi_i\rangle+\langle\psi_i|\Sigma(E_i)|\psi_i\rangle
 \end{equation}
where $\psi_i$ and $\epsilon_i$ are the $i$-th Kohn-Sham orbital and its Kohn-Sham energy, $V_\mathrm{xc}$
is the DFT exchange-correlation potential and $\Sigma$ is the self-energy operator in the $G_0W_0$ approximation.
As the calculated $G_0W_0$ QP energy levels depend upon the choice of the exchange-correlation functional in the starting DFT calculation,
we calculated QP energy levels from both GGA-PBE and the local-density approximation (LDA) exchange-correlation functional (LDA) in the Perdew-Zunger form (PZ) \cite{PZ81}. For the two sets of calculations
we used the same PBE structural parameters.
It is worth noticing that QP energy levels of occupied orbitals correspond to vertical ionization potentials (VIPs) 
while those of unoccupied ones correspond to vertical electron affinities (VEAs), which are closely related to the open-circuit voltage discussed later.

In order to remove artificial periodic image interactions, we employed truncated Coulomb potentials 
with a spherical radius cutoff of 15.9, 15.9, 9.6, and 23.8 {\AA} for C$_{60}$, C$_{70}$, [C$_{60}$]PCBM, and bis-[C$_{60}$]PCBM, respectively. 
{For bis-[C$_{60}$]PCBM, we checked the convergence with respect to the radius cutoff by performing {\it GW} calculations starting from LDA and GGA 
in a smaller cell of  32.0$\times$19.2$\times$19.2 \AA$^3$ with a smaller radius cutoff of 9.6~\AA. We considered the states close to the highest-occupied molecular orbital (HOMO) and the lowest-unoccupied molecular orbital (LUMO), and observed 
only an average increase of VIPs and VEAs of 0.03 (0.03) eV for LDA (GGA). This corroborates the choice of a small (9.6 \AA) radius cutoff for [C$_{60}$]PCBM 
and for addressing the entire electronic DOS of bis-[C$_{60}$]PCBM}.
The {\it GW} calculations were performed with the {\it GWL} code described in  Refs.~\onlinecite{UmariSB09, UmariSB10, UmariQMSGB11, QianUM11}. This approach permits calculations for relatively large atomic structures by expanding the polarizability operators on optimal basis sets. 

{An optimal basis for representing polarizability operator is given by the most important 
({\it i.e.}, corresponding to the largest eigenvalues) eigenvectors of an easy-to-calculate  average polarizability operator.
This is defined by the products of occupied orbitals with a set of plane-waves which were first projected onto the
conduction manifold and then orthonormalized. We indicate with  $E^*$ the energy cutoff defining such plane-waves
basis set and with  $q^*$ the threshold controlling the final number of elements in our polarizability basis.
The accuracy of the final {\it GW} levels depends on the interplay between  $E^*$ and  $q^*$.
Larger values for $E^*$ yield more accurate results although requiring smaller  $q^*$ and hence  larger final basis sets.
Smaller values for $E^*$ yield less accurate results, but require a smaller number of final basis sets permitting to afford larger model structures.}

We used  a parameter $E^*$ of 68.0 eV and a threshold $q^*$ of 1.0 a.u., resulting in 4381, 5645, and 7005 optimal polarizability basis elements for C$_{60}$, C$_{70}$, and PCBM, respectively. The slow convergence of the sum-over-empty-states in conventional {\it GW} implementations are completely avoided  by using a Lanczos chain approach. Here chains of 20 Lanczos steps were applied for the polarizability operators and chains of 120 Lanczos steps were used for the self-energy expectation values. Those were first evaluated on the imaginary energy axis and then analytically continued onto the real one by fitting with a two-pole expansion. Fitting with a three-pole expansion or changing the energy range in the fitting yielded differences of less than $50$ meV for the energy levels of frontier orbitals. We estimate a computational accuracy of $0.1$ eV for the calculated QP levels with respect to the vacuum level.

\section{{\it GW} Quasiparticle Energy Levels}
\label{SEC:RESULTS}

\begin{table*}[thpb]
  \caption{Low-lying vertical ionization potentials and vertical electron affinities of C$_{60}$ and their corresponding orbitals (unit: eV), where we also list
the $G_0W_0$@LDA data of Ref.~\onlinecite{BlaseAO11}, and $G_0W_0$@LDA data of Ref.~\onlinecite{TiagoKHR08}.}
  \label{TABLE1} \setlength{\extrarowheight}{3pt}
  \begin{center}
    \begin{tabular}{
        >{\centering}p{0.1\columnwidth}
        >{\centering}p{0.15\columnwidth}|
        >{\centering}p{0.14\columnwidth}
        >{\centering}p{0.14\columnwidth}|
        >{\centering}p{0.14\columnwidth}
        >{\centering}p{0.14\columnwidth}
        >{\centering}p{0.14\columnwidth}
        >{\centering}p{0.14\columnwidth}|
        >{\centering}p{0.20\columnwidth}}
      \hline
      \hline
       & Orbital & \multicolumn{2}{c}{DFT} & \multicolumn{4}{c|}{$G_0W_0$} & Expt. \cite{WangDW99, LichtenbergerNRHL91} \tabularnewline
      \hline 
      & & \multicolumn{1}{c}{LDA} & GGA & LDA & LDA \cite{BlaseAO11} & LDA \cite{TiagoKHR08}& GGA  & \tabularnewline[+4pt]
      \multirow{3}{*}{VEA}
      & $t_{2u}$ & 2.32 & 2.17  &  0.84  & & &0.70  & \tabularnewline
      & $t_{1g}$ & 3.21 & 3.07  & 1.81  & & &1.65  & \tabularnewline
      & $t_{1u}$ & 4.37 & 4.16  & 3.04  & 2.84 & 3.87 & 2.82& 2.69 \tabularnewline[3pt]
      \hline
      \multirow{10}{*}{VIP}
      & $h_u$ & 6.05  & 5.83  & 7.68 & 7.28 &  8.22 & 7.37 &  7.6 \tabularnewline
      & $g_g$ & 7.28  & 6.99  & 9.01 & & 9.33 &8.68 & \multirow{2}{*}{8.95} \tabularnewline
      & $h_g$ & 7.40  & 7.10  & 9.07 & & 9.42 &8.69 &   \tabularnewline
      & $h_u$ & 8.89  & 8.63  & 11.32 & & 11.93& 10.94 &   10.82 \tabularnewline
      & $g_u$ & 9.03  & 8.66  & 10.65 & & 11.00 & 10.45 &  $\wr$ \tabularnewline
      & $t_{2u}$& 9.60  & 9.21  &11.23 & & 11.46 & 10.91&   11.59 \tabularnewline
      & $h_g$ &   9.21  & 8.93  & 11.67 & & 12.19& 11.25 &  12.43 \tabularnewline
      & $g_u$ &  10.23 & 9.95  & 12.76 & & 13.23 &12.35 &  \multirow{2}{*}{$\wr$} \tabularnewline
      & $t_{1g}$& 10.75 & 10.44  & 13.26 & & 13.60 &12.87  &  \tabularnewline
      & $h_g$ &  10.95 & 10.50  & 12.59 & & 12.74 &12.33 &  13.82 \tabularnewline[3pt]
      \hline
      $E_{\rm gap}$ & & 1.68 & 1.66  & 4.64 & 4.44 & 4.35 &4.55 & 4.91 \tabularnewline[3pt]
      \hline
      \hline
    \end{tabular}
  \end{center}
\end{table*}

\begin{table}[thpb]
  \caption{Low-lying vertical ionization potentials and vertical electron affinities of C$_{70}$ and their corresponding excited states. (unit: eV)}
  \label{TABLE2} \setlength{\extrarowheight}{3pt}
  \begin{center}
    \begin{tabular}{
        >{\centering}p{0.1\columnwidth}
        >{\centering}p{0.14\columnwidth}|
        >{\centering}p{0.11\columnwidth}
        >{\centering}p{0.11\columnwidth}|
        >{\centering}p{0.11\columnwidth}
        >{\centering}p{0.11\columnwidth}|
        >{\centering}p{0.20\columnwidth}}
      \hline
      \hline 
       & Orbital & \multicolumn{2}{c}{DFT}  & \multicolumn{2}{c|}{$G_0W_0$} & Expt. \cite{WangWHKW06, LichtenbergerRG92} \tabularnewline
      \hline 
      & & \multicolumn{1}{c}{LDA} & GGA  & LDA & GGA & \tabularnewline
      [+4pt]
      \multirow{2}{*}{VEA}
      & $a_{1}''$ & 4.17 & 4.00  & 3.03 & 2.81 & \tabularnewline
      & $e_{1}''$ & 4.30 & 4.10  & 3.13 & 2.91 & 2.765 \tabularnewline[3pt]
      \hline
      \multirow{11}{*}{VIP}
      & $a_{2}''$ & 6.14 & 5.87  & 7.54 & 7.21 & 7.47 \tabularnewline
      & $e_{1}''$ & 6.06 & 5.82  & 7.59 & 7.29 & 7.47 \tabularnewline
      & $a_{2}'$  & 6.29 & 6.05  & 7.88 & 7.53 & 7.68 \tabularnewline
      & $e_{2}'$  & 6.42 & 6.25  & 8.12 & 7.79 & 7.96 \tabularnewline
      & $e_{2}''$ & 6.50 & 6.16  & 8.14 & 7.79 & 8.12 \tabularnewline
      & $e_{1}'$  & 6.78 & 6.50  & 8.44 & 8.10 & 8.43 \tabularnewline
      & $e_{1}'$  & 7.53 & 7.22  & 9.27 & 8.87 & 9.04 \tabularnewline
      & $e_{2}''$ & 7.94 & 7.61  & 9.75 & 9.34 & 9.28 \tabularnewline
      & $e_{1}''$ & 8.07 & 7.73  & 9.93 & 9.47 & 9.60 \tabularnewline
      & $e_{2}'$  & 8.15 & 7.81  & 9.95 & 9.54 & 9.60 \tabularnewline
      & $a_{1}'$  & 8.57 & 8.21  & 10.34 & 9.88 & 9.84 \tabularnewline[3pt]
      \hline
      $E_{\rm gap}$ &  & 1.84 & 1.77  & 4.42 & 4.30 & 4.71 \tabularnewline[3pt]
      \hline
      \hline 
    \end{tabular}
  \end{center}
\end{table}

\begin{table}[thpb]
  \caption{Low-lying vertical ionization potentials and vertical electron affinities of [C$_{60}$]PCBM and their corresponding excited states. (unit: eV)}
  \label{TABLE3} \setlength{\extrarowheight}{3pt}
  \begin{center}
    \begin{tabular}{
        >{\centering}p{0.1\columnwidth}
        >{\centering}p{0.16\columnwidth}|
        >{\centering}p{0.11\columnwidth}
        >{\centering}p{0.11\columnwidth}|
        >{\centering}p{0.11\columnwidth}
        >{\centering}p{0.11\columnwidth}|
        >{\centering}p{0.18\columnwidth}}
      \hline
      \hline
       & Orbital & \multicolumn{2}{c}{DFT} & \multicolumn{2}{c|}{$G_0W_0$} & Expt. \cite{AkaikeKYTNSOS08} \tabularnewline
      \hline 
      & & \multicolumn{1}{c}{LDA} & GGA  & LDA & GGA & \tabularnewline
      [+4pt]
      \multirow{6}{*}{VEA}
      &  \multirow{3}{*}{C$_{60}$($t_{1g}$)} & 2.61 & 2.45  & 1.42 & 1.18  & \tabularnewline
      &  
         & 2.84 & 2.67   & 1.66 & 1.44  & \tabularnewline
      &  & 2.87 & 2.70   & 1.72 & 1.49  & \tabularnewline[8pt]
      & \multirow{3}{*}{C$_{60}$($t_{1u}$)} & 3.77 & 3.55   & 2.67 & 2.40  & \tabularnewline
      & 
         & 4.00 & 3.77   & 2.92 & 2.64  & \tabularnewline
      &  & 4.03 & 3.80   & 2.95 & 2.69  & \tabularnewline[3pt]
      \hline
      & \multirow{5}{*}{C$_{60}$($h_{u}$)}  & 5.53 & 5.30  & 7.01 & 6.72 &  7.17 \tabularnewline
      & & 5.67 & 5.43 & 7.17 & 6.88 &  \tabularnewline
      VIP & 
         & 5.67 & 5.43  & 7.14 & 6.85 & \tabularnewline
      &  & 5.71 & 5.47  & 7.21 & 6.91 &  \tabularnewline
      &  & 5.89 & 5.63  & 7.38 & 7.08 & \tabularnewline[3pt]
      \hline
      $E_{\rm gap}$ & & 1.50 & 1.50  & 4.06 & 4.03 \tabularnewline[3pt]
      \hline
      \hline 
    \end{tabular}
  \end{center}
\end{table}

\begin{table}[thpb]
  \caption{Low-lying vertical ionization potentials and vertical electron affinities of metastable [C$_{60}$]PCBM and their corresponding excited states. (unit: eV)}
  \label{TABLE4} \setlength{\extrarowheight}{3pt}
  \begin{center}
    \begin{tabular}{
        >{\centering}p{0.1\columnwidth}
        >{\centering}p{0.16\columnwidth}|
        >{\centering}p{0.11\columnwidth}
        >{\centering}p{0.11\columnwidth}|
        >{\centering}p{0.11\columnwidth}
        >{\centering}p{0.11\columnwidth}|
        >{\centering}p{0.18\columnwidth}}
      \hline
      \hline
       & Orbital & \multicolumn{2}{c}{DFT}  & \multicolumn{2}{c|}{$G_0W_0$} & Expt. \cite{AkaikeKYTNSOS08} \tabularnewline
      \hline 
      & & \multicolumn{1}{c}{LDA} & GGA  & LDA & GGA & \tabularnewline
      [+4pt]
      \multirow{6}{*}{VEA}
      &  \multirow{3}{*}{C$_{60}$($t_{1g}$)}  & 2.85 & 2.68  & 1.67 & 1.41 & \tabularnewline
      & 
         & 3.08 & 2.90  & 1.90 & 1.67 & \tabularnewline
      &  & 3.11 & 2.93  & 1.97 & 1.73 & \tabularnewline[8pt]
      & \multirow{3}{*}{C$_{60}$($t_{1u}$)}  & 4.00 & 3.78  & 2.92 & 2.63 & \tabularnewline
      &
         & 4.24 & 3.99  & 3.15 & 2.85 & \tabularnewline
      &  & 4.27 & 4.03  & 3.18 & 2.94 & \tabularnewline[3pt]
      \hline
      & \multirow{5}{*}{C$_{60}$($h_{u}$)} & 5.75 & 5.51  & 7.24 & 6.92 &  7.17 \tabularnewline
      &  & 5.91 & 5.66  & 7.41 & 7.10 &  \tabularnewline
      VIP &
         & 5.91 & 5.66  & 7.36 & 7.05 & \tabularnewline
      &  & 5.94 & 5.69  & 7.44 & 7.11 &  \tabularnewline
      &  & 6.11 & 5.86  & 7.61 & 7.30 & \tabularnewline[3pt]
      \hline
      $E_{\rm gap}$ & & 1.48 & 1.48  & 4.06 & 3.98 \tabularnewline[3pt]
      \hline
      \hline 
    \end{tabular}
  \end{center}
\end{table}

\begin{table}[thpb]
  \caption{Low-lying vertical ionization potentials and vertical electron affinities of bis-[C$_{60}$]PCBM and their corresponding excited states. (unit: eV)}
  \label{TABLE5} \setlength{\extrarowheight}{3pt}
  \begin{center}
    \begin{tabular}{
        >{\centering}p{0.1\columnwidth}
        >{\centering}p{0.16\columnwidth}|
        >{\centering}p{0.11\columnwidth}
        >{\centering}p{0.11\columnwidth}|
        >{\centering}p{0.11\columnwidth}
        >{\centering}p{0.11\columnwidth}|
        >{\centering}p{0.18\columnwidth}}
      \hline
      \hline
       & Orbital & \multicolumn{2}{c}{DFT}  & \multicolumn{2}{c|}{$G_0W_0$} & Expt. (N.A.)\tabularnewline
      \hline 
      & & \multicolumn{1}{c}{LDA} & GGA  & LDA & GGA & \tabularnewline
      [+4pt]
      \multirow{6}{*}{VEA}
      & \multirow{3}{*}{C$_{60}$($t_{1g}$)}   & 2.34 & 2.19  & 1.23 & 1.05  & \tabularnewline
      &
         & 2.71 & 2.53   & 1.62 & 1.39  & \tabularnewline
      &  & 2.75 & 2.58   & 1.70 & 1.48  & \tabularnewline[8pt]
      & \multirow{3}{*}{C$_{60}$($t_{1u}$)}   
      & 3.29 & 3.08   & 2.27 & 2.02  & \tabularnewline
      & 
         & 3.86 & 3.62   & 2.87 & 2.59  & \tabularnewline
      &  & 3.91 & 3.68   & 2.94 & 2.68  & \tabularnewline[3pt]
      \hline
      & \multirow{5}{*}{C$_{60}$($h_{u}$)} 
      & 5.35 & 5.11  & 6.72 & 6.42 &  \tabularnewline
      &  & 5.52 & 5.28  & 6.91 & 6.59 &  \tabularnewline
      VIP &
         & 5.53 & 5.29  & 6.92 & 6.60 & \tabularnewline
      &  & 5.61 & 5.36  & 7.04 & 6.72 & \tabularnewline
      &  & 6.00 & 5.75  & 7.48 & 7.16 & \tabularnewline[3pt]
      \hline
      $E_{\rm gap}$ & & 1.44 & 1.43  & 3.78 & 3.74 \tabularnewline[3pt]
      \hline
      \hline 
    \end{tabular}
  \end{center}
\end{table}

In Tables~\ref{TABLE1} and ~\ref{TABLE2} we compare the calculated VIPs and VEAs of C$_{60}$ and C$_{70}$ with experimental data. In these two cases, it has been possible to resolve single lines in the experimental photoemission spectra other than the first VEA and VIP. Thanks to the lower symmetry of the molecule, the lines corresponding to all the orbitals have  been obtained for C$_{70}$. For simplicity we indicate {$G_0W_0$}@LDA ({$G_0W_0$}@GGA) the results for $G_0W_0$ calculated starting from LDA (GGA). We see that the first VIP calculated by {$G_0W_0$}@LDA is larger than that from {$G_0W_0$}@GGA. {$G_0W_0$}@LDA overestimates the first VIP up to $0.2$ eV  for C$_{70}$ while {$G_0W_0$}@GGA underestimates it by roughly the same amount. A different trend is found for the first VEAs: both {$G_0W_0$}@LDA and {$G_0W_0$}@GGA overestimate the experimental value by about $0.4$ eV and $0.2$ eV, respectively. This yields similar values in the two approximations for the HOMO-LUMO gaps, which are $0.3$ eV lower than experiment. Furthermore, the relative error between higher VIPs and experiment is quite stable and similar to that of the first VIPs, suggesting a good description of the photoemission spectra. Such behavior is not found for DFT calculations, leading to both absolute and relative discrepancies in VIPs and VEAs. In addition, results for ground-state [C$_{60}$]PCBM, metastable [C$_{60}$]PCBM, and bis-[C$_{60}$]PCBM are reported in Tables~\ref{TABLE3}, \ref{TABLE4}, and \ref{TABLE5}. In this case, only the experimental VIP for [C$_{60}$]PCBM is available, exhibiting the same degree of accuracy as that found for C$_{60}$ and C$_{70}$. 

{As listed in Table~\ref{TABLE1} for C$_{60}$, the quantitative description of quasiparticle energies at the $G_0W_0$ level
 can be improved by solving the bottleneck of the sum-over-empty-states through the efficient Lanczos approach \cite{UmariSB09,
  UmariSB10, StenuitCPFPGU10, UmariQMSGB11, QianUM11}. Indeed, our {\it $G_0W_0$}@LDA results improve upon those of Ref.~\onlinecite{TiagoKHR08} where a static remainder
is used which accounts for  infinite sums over empty states \cite{Tiago2006}. Our {\it $G_0W_0$}@LDA results for VEA and VIP are in a closer agreement
with those from Ref.~\onlinecite{BlaseAO11} where Gaussian basis sets were used allowing for summing over all the available empty states.
Our {\it $G_0W_0$}@LDA VIP and VEA levels are found in the middle between those of Ref.~\onlinecite{TiagoKHR08} and those 
of Ref.~\onlinecite{BlaseAO11}.  
} 
The present results are also in good agreement with the predictions from Koopmans' compliant functionals \cite{Koopmans10, Koopmans14a}, where for e.g. C$_{60}$ {using the PBE functional}  the VIP and VEA are 7.42 eV and 2.82 eV \cite{Koopmans14b}, against the present {$G_0W_0$}@GGA  results of 7.37 eV and 2.82 eV, respectively. 

\begin{figure}[thpb]\centering
  \includegraphics[width=0.95\columnwidth]{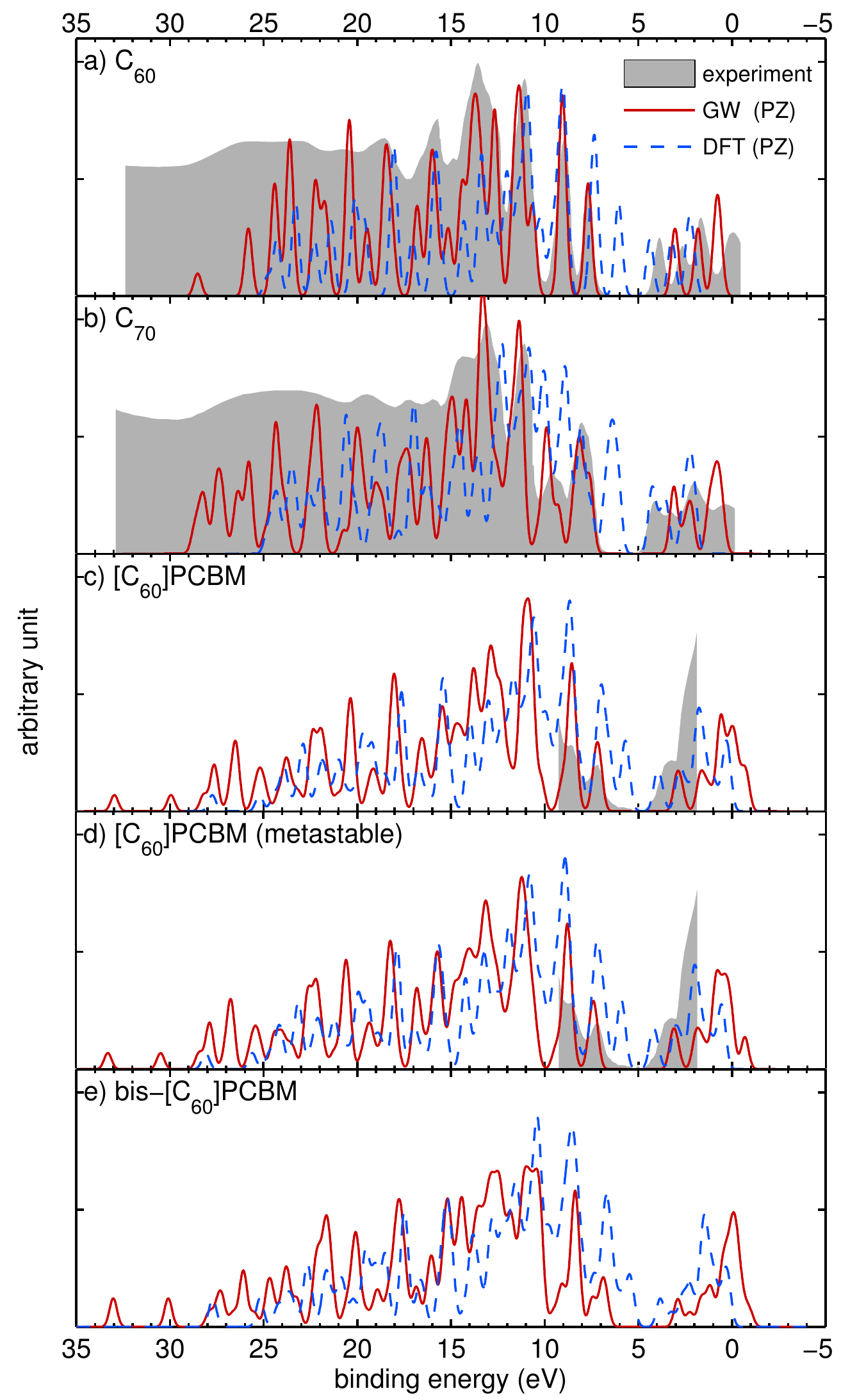}
  \caption{Density of states from {\it $G_0W_0$}@LDA  and DFT calculations
    compared with experimental photoemission and inverse photoemission
    spectra for (a) C$_{60}$, (b) C$_{70}$, (c) [C$_{60}$]PCBM, (d)
    metastable-[C$_{60}$]PCBM, and (e) bis-[C$_{60}$]PCBM. 
    Experimental data of PES and IPES were adopted from
    Refs. \onlinecite{BenningPOCJSKWFS92},
    \onlinecite{BenningPOCJSKWFS92}, \onlinecite{AkaikeKYTNSOS08} for
    C$_{60}$, C$_{70}$, and [C$_{60}$]PCBM in their thin-films, respectively.}
  \label{FIG:FULLERENE_PES_IPES_LDA}
\end{figure}

\begin{figure}[thpb]\centering
  \includegraphics[width=0.95\columnwidth]{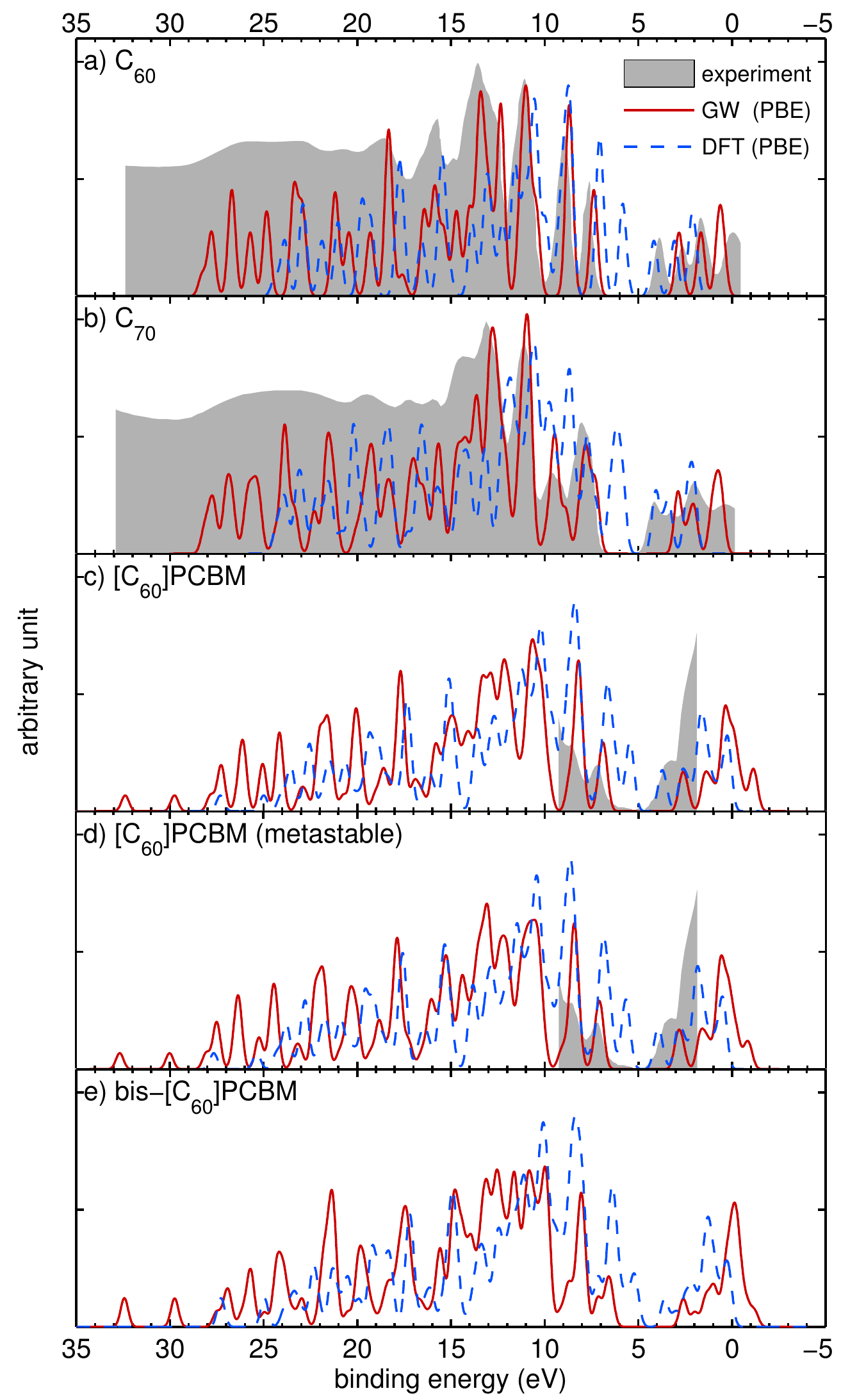}
  \caption{Density of states from {\it $G_0W_0$}@GGA  and DFT calculations
    compared with experimental photoemission and inverse photoemission
    spectra for (a) C$_{60}$, (b) C$_{70}$, (c) [C$_{60}$]PCBM, (d)
    metastable-[C$_{60}$]PCBM, and (e) bis-[C$_{60}$]PCBM. 
    Experimental data of PES and IPES were adopted from
    Refs. \onlinecite{BenningPOCJSKWFS92},
    \onlinecite{BenningPOCJSKWFS92}, \onlinecite{AkaikeKYTNSOS08} for
    C$_{60}$, C$_{70}$, and [C$_{60}$]PCBM in their thin-films, respectively.}
  \label{FIG:FULLERENE_PES_IPES_GGA}
\end{figure}

We display in Figs.~\ref{FIG:FULLERENE_PES_IPES_LDA} and \ref{FIG:FULLERENE_PES_IPES_GGA} the electronic density of states (DOS) of the five electron-acceptor molecules, calculated with {\it $G_0W_0$}@LDA and {\it $G_0W_0$}@GGA, respectively, and compare them with direct and inverse photoemission data without peak alignment, albeit neglecting any oscillator strength effect in the calculation.  We can see that both {$G_0W_0$}@LDA and {$G_0W_0$}@GGA give a good description of the photoemission spectra.  It is worth noting that the lower parts of the conduction DOSs are mainly due to bound orbitals, and they are well described by isolated molecules as can be seen from their excellent agreement with inverse photoemission. In addition, the electronic band gap of thin films is strongly reduced in the bulk with respect to the gas-phase, due to the large dielectric screening \cite{Marsili2014}. 

{
It is worth to mention that [C$_{60}$]PCBM and metastable [C$_{60}$]PCBM exhibit a very similar DOS at both the DFT 
and the $G_0W_0$ level with the metastable [C$_{60}$]PCBM shifted by about 0.25 eV towards lower energies. 
This should be ascribed  to the analogous geometry of the two systems accompanied by an increase of the
nuclei electrostatic potential for metastable [C$_{60}$]PCBM in which the functional group is closer to the C$_{60}$ part than that of ground-state [C$_{60}$]PCBM.}

The overall good agreement together with the fact that the main difference with respect to experiments can be  described as a rigid shift of the HOMO-LUMO gap by less than about $0.3$ eV supports the quality of {\it GW} results for further investigations of variations in ${\rm V}_\mathrm{oc}$. 


\section{Role of self-consistency and starting DFT flavors}
\label{secself}
{There are two major factors responsible for the difference between {\it $G_0W_0$} energy levels and experimental values, including the starting DFT flavor and the non-self-consistency of the {\it $G_0W_0$} scheme. We address these effects by carrying out self-consistent {\it GW} calculations. In this way, we can single out the effects due to different DFT orbitals as 
we retain the diagonal approximation. We implemented a simplified self-consistent {\it GW} scheme. At each iteration, we rigidly 
displace the energy of the entire valence and the energy of the conduction manifold. In practice, after the first  {\it $G_0W_0$} calculation, we start an iterative series of analogous {\it GW} calculations in which all the starting DFT energies of valence states
are shifted by $s_v$ and all the starting DFT energies of conduction states are shifted by  $s_c$.
This can be achieved by avoiding the sum-over-empty-orbitals through the application of the following operator:
}
\begin{equation}
\hat{S}=(1+s_c)\mathbb{I}+(s_v-s_c)\hat{P}_v,
\end{equation}
{
where $\hat{P}_v$ is the projector over the DFT valence manifold and  $s_v$ and  $s_c$  
are chosen in order to align HOMO and LUMO levels with the corresponding  {\it GW} values of the
previous iteration. 
At variance with previous implementations using a rigid scissor \cite{Marina2014}, it is important not only 
to update the HOMO-LUMO band gap but also their actual levels as we are interested in the absolute values of the energy levels.
For the sake of simplicity, we focus on the VIP and VEA of C$_{60}$ starting from LDA and GGA. The results are reported in Table~\ref{tabself}.
}

\begin{table}[thpb]
  \caption{{\it GW}  values for VIP and VEA of C$_{60}$ calculated using the self-consistent
   scheme of Sec. \ref{secself}. {\it $G_nW_n$} indicates the $n$-th iteration. Units: eV}
  \label{tabself} \setlength{\extrarowheight}{3pt}
  \begin{center}
  \begin{tabular}{l | ccccc}
      \hline
      \hline
        & &{\it $G_0W_0$} & {\it $G_1W_1$} & {\it $G_2W_2$} & Expt. \cite{WangDW99, LichtenbergerNRHL91} \\
\hline
  & VEA &    3.04  &   3.12  &   3.12 & 2.69 \\
LDA &   &           &         & &\\
  & VIP &    7.69  &   8.09  &   8.12 & 7.6 \\
\hline
  & VEA &   2.82   &  2.89   &  2.90  & 2.69\\
GGA & &           &       &&\\
 & VIP &   7.37   &  7.80  &  7.84 & 7.6\\
\hline
\hline
  \end{tabular}
  \end{center}
\end{table}
{
We note that VIP and VEA have already converged within a few tens of meV at the second iteration after the first {\it $G_0W_0$} run.
In contrast with {\it $G_0W_0$}, self-consistent {\it GW}@GGA gives both VIP and VEA in good agreement (within $\sim0.2$ eV) with experiment while self-consistent {\it GW}@LDA overestimates VIP and VEA by more than $0.4$ eV. Moreover,  regardless of the LDA or GGA flavor, the  {\it GW} VEA increases by $\sim0.1$ eV going from {\it $G_0W_0$} to self-consistent {\it GW}, while VIP increases by $\sim0.4$ eV. This permits to ascribe the differences between {\it GW}@GGA  and {\it GW}@LDA to the different quality of the DFT orbitals, {\it i.e.}, the GGA ones being more accurate.
}
\section{${\rm V}_{\mathrm oc}$ and the role of functional side group of PCBM}
\label{SEC:COMPARISON}

In actual devices the experimental open-circuit voltage depends not only on the conditions during the measurements (e.g. illumination) but also on the geometry of the cells and the morphology of their constituents. Here, we address only the maximum attainable ${\rm V}_{\mathrm oc}$ which, in a single particle picture, is determined by the energy difference between the LUMO level of the electron acceptor and the HOMO level of the electron donor while neglecting structural relaxations accompanying charged excitations. The former corresponds to minus the first-VEA of the acceptor and the latter to minus the first-VIP of the donor. Consequently, ${\rm V}_{\mathrm oc} = $ first-VIP (donor) $-$ first-VEA (acceptor). Here, we focus only on the differences in the maximum attainable open-circuit voltage, $\Delta{\rm V}_{\mathrm oc}  ({\mathrm A} \vert {\mathrm B})$,  between electron acceptor A and B. That is, $\Delta{\rm V}_{\mathrm oc} ({\mathrm A} \vert {\mathrm B}) = {\rm V}_{\mathrm oc}({\mathrm A}) - {\rm V}_{\mathrm oc}({\mathrm B})$. Therefore, the comparison with experimental ${\rm V}_{\mathrm oc}$'s would be meaningful only for similar devices.

{
Quasiparticle energy levels may vary from the limit of an isolated molecule to that of a bulk. In particular, the HOMO-LUMO gap can be significantly  reduced \cite{Marsili2014}. However, in weakly-bonded crystalline or disordered bulks  the main effect comes from static dielectric screening. Assuming similar static dielectric properties we can expect  similar  changes in the LUMO levels. It is also worth to note that the realistic modelling of such electron acceptor layers is quite a demanding task for structural optimization using DFT and electronic structure calculations using the {\it GW} approximation. Therefore, it is important to find ways to estimate $V_{oc}$ addressing the limit of isolated systems. 
}

\begin{figure}[thpb]\centering
  \includegraphics[width=0.9\columnwidth]{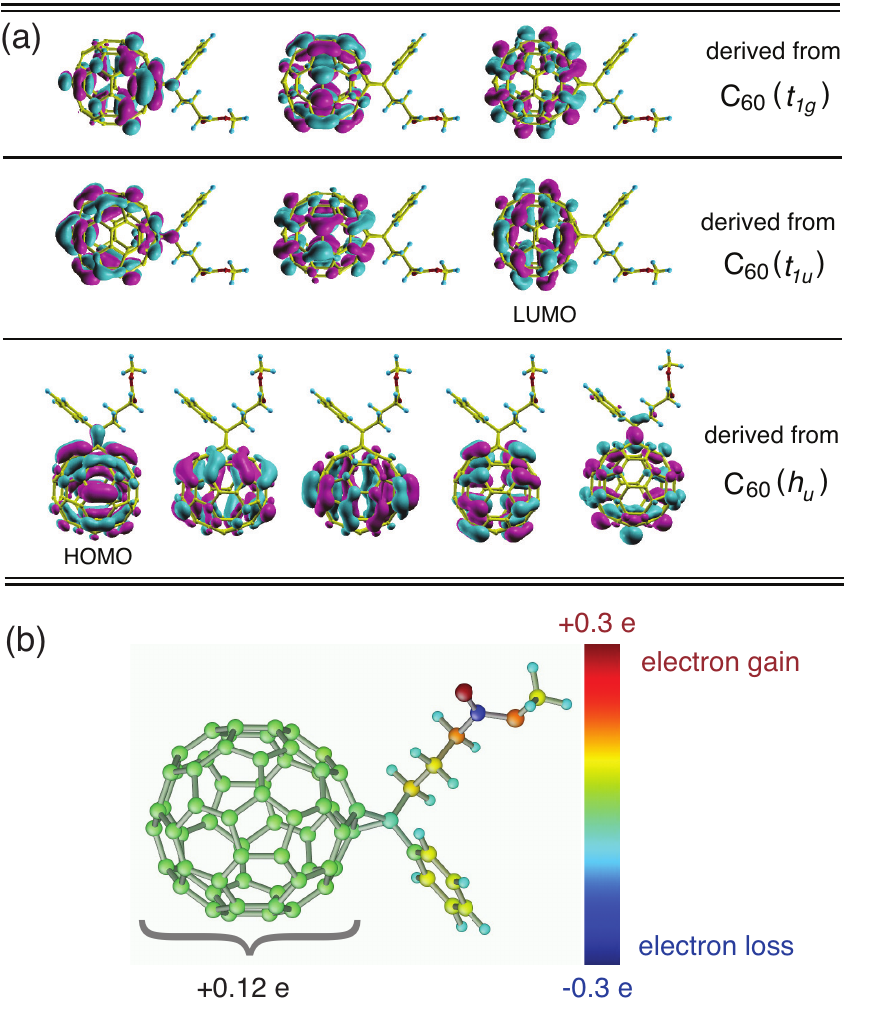}
  \caption{(a) Quasiparticle states close to HOMO and LUMO in the ground-state [C$_{60}$]PCBM. States shown in the first, second, and third rows are derived from fullerene C$_{\rm 60}$'s $t_{1g}$, $t_{1u}$, and $h_{u}$, respectively. Purple (cyan) indicates positive (negative) value of wave functions. (b) L\"{o}wdin atomic charge analysis of [C$_{60}$]PCBM. Red (blue) indicates electron gain (loss). About 0.12 electron was transferred from the side group to the fullerene cage.}
  \label{FIG:PCBM_ORBITAL_LOWDIN}
\end{figure}

\begin{table*}[thpb]
\caption{Theoretical and experimental differences in the maximum attainable open-circuit voltage, $\Delta{\rm V}_{\mathrm oc}  ({\mathrm A} \vert {\mathrm B})$,  between electron acceptor A and B (unit: eV). $\Delta{\rm V}_{\mathrm oc} ({\mathrm A} \vert {\mathrm B}) = {\rm V}_{\mathrm oc}({\mathrm A}) - {\rm V}_{\mathrm oc}({\mathrm B})$. We indicate with $^{*}$ the
metastable structure of [C$_{60}$]PCBM.} 
\label{TABLE6} \setlength{\extrarowheight}{5pt}
\begin{threeparttable}
\begin{tabular}{
        >{\centering}p{0.3\columnwidth}
        >{\centering}p{0.02\columnwidth}
        >{\centering}p{0.3\columnwidth}|
        >{\centering}p{0.16\columnwidth}|
        >{\centering}p{0.16\columnwidth}|
        >{\centering}p{0.24\columnwidth}|
        >{\centering}p{0.24\columnwidth}|
        >{\centering}p{0.24\columnwidth}}
\hline
\hline
\multicolumn{3}{c|}{$\Delta{\rm V}_{\mathrm oc}$(A$\vert$B) (eV)}  & LDA    & GGA    & $G_0W_0$@LDA & $G_0W_0$@GGA & Expt. \tabularnewline[3pt]
\hline
\multicolumn{1}{r}{C$_{70}$} &   $\vert$  &  \multicolumn{1}{l|}{C$_{60}$}   &  0.08  &  0.06  &  -0.09   & -0.09  &  0\tnote{a} \tabularnewline[3pt]
\multicolumn{1}{r}{[C$_{60}$]PCBM} &  $\vert$ &  \multicolumn{1}{l|}{C$_{60}$}   &  0.35  &  0.37  &   0.09   &  0.13  & 0.15\tnote{b} \tabularnewline[3pt]
\multicolumn{1}{r}{[C$_{60}$]PCBM$^{*}$}  &  $\vert$ &  \multicolumn{1}{l|}{C$_{60}$} &  0.11  &  0.13  &  -0.15   & -0.12  & \tabularnewline[3pt]
\multicolumn{1}{r}{bis-[C$_{60}$]PCBM} &  $\vert$ & \multicolumn{1}{l|}{[C$_{60}$]PCBM}  &  0.12  &  0.12  &   0.01   &  0.01  & 0.12\tnote{c}   \tabularnewline[3pt]
\hline
\end{tabular}
        \begin{tablenotes}
            \item[a] Ref.~\onlinecite{Pfuetzner2009}
            \item[b] Ref.~\onlinecite{BrabecCMSFRSH01}
            \item[c] Averaged value from 0.15 eV of Ref.~\onlinecite{LenesWKVHB08},  and 0.23 eV, 0.09 eV, 0.02 eV from Ref.~\onlinecite{Tromholt2009}.
        \end{tablenotes}
\end{threeparttable}
\end{table*}

We report in Table~\ref{TABLE6} the differences in the maximum attainable open-circuit voltages, $\Delta{\rm V}_{\mathrm oc} ({\mathrm A} \vert {\mathrm B})$.
{ The {\it GW} calculations can fairly reproduce the experimental $\Delta{\rm V}_{\mathrm oc}$ for  [C$_{60}$]PCBM taking care of the large range of experimental values.}
The difference in LUMO energy between C$_{60}$ and  [C$_{60}$]PCBM is small, which can be traced back to the fact that the lowest unoccupied orbitals of [C$_{60}$]PCBM are derived from the C$_{60}$ three-fold degenerate LUMO. Similarly, the highest occupied orbitals of  [C$_{60}$]PCBM are derived from the C$_{60}$ five-fold degenerate HOMO. These orbitals are displayed in Fig.~\ref{FIG:PCBM_ORBITAL_LOWDIN}(a) where one can appreciate their localization on the C$_{60}$ group. 
As highlighted from the L\"{o}wdin charge analysis \cite{Qian08} displayed in Fig.~\ref{FIG:PCBM_ORBITAL_LOWDIN}(b), 
a  moderate charge transfer of 0.12~{\it e} occurs towards the fullerene cage. 
{
Going from  C$_{60}$ to  [C$_{60}$]PCBM, we observe an upshift of the HOMO level and a relatively smaller upshift of the LUMO level, leading to a smaller HOMO-LUMO gap.
An analogous behavior  is registered from [C$_{60}$]PCBM to bis-[C$_{60}$]PCBM where a larger charge transfer is expected. 
}

In contrast, for the metastable [C$_{60}$]PCBM  configuration we predict a lower ${\rm V}_{\mathrm oc}$ than for C$_{60}$. The {fair} agreement between {\it GW} 
and experiment for [C$_{60}$]PCBM and bis-[C$_{60}$]PCBM is not observed for C$_{70}$. In this case, both DFT and {\it GW} indicate a smaller ${\rm V}_{\mathrm oc}$ while experimental ${\rm V}_{\mathrm oc}$ shows almost no variation. This must be ascribed to using an isolated molecule approximation, since DFT and {\it GW} reproduce well the VEA of C$_{60}$ and C$_{70}$ molecules and their order. More accurate results would require the modeling of bulk materials and interfaces.

\section{Conclusions}
\label{SEC:SUMMARY}

{\it GW} approaches are very promising for determining the main physical characteristics of polymer solar cells. The isolated  molecular limit adopted in this work gives overall good results, in particular for the [C$_{60}$]PCBM and bis-[C$_{60}$]PCBM. 
These results lead us to expect that more accurate agreement may be obtained by direct modeling of bulk photovoltaic materials and interfaces using the {\it GW}-Lanczos approach, which is particularly suitable for large systems. Finally, the comparison between ground-state [C$_{60}$]PCBM and metastable [C$_{60}$]PCBM demonstrates that the intrinsic morphological difference can have significant effects on the open-circuit voltage, providing a theoretical confirmation of morphology as an important factor in organic photovoltaics.

\begin{acknowledgments}
This work was supported by the Department of Energy SciDAC program on Quantum Simulations of Materials and Nanostructures (DE-FC02-06ER25794) and Eni S.p.A. under the Eni-MIT Alliance Solar Frontiers Program. Calculations were performed on the CINECA HPC-facility thanks to Prace allocation 2011050812 and Iscra allocation HP10AXPUBZ.
\end{acknowledgments}

%

\end{document}